\documentclass[prl,twocolumn,showpacs,superscriptaddress,amsmath,amssymb]{revtex4}

\usepackage{graphicx}
\usepackage{dcolumn}
\usepackage{bm}

\begin{document}

\title{Spin Order and Excitations in Triangular Antiferromagnet La$_2$Ca$_2$MnO$_7$}

\author{Wei Bao}
\email{wbao@ruc.edu.cn}
\affiliation{Department of Physics, Renmin University of China, Beijing 100872, China}
\author{Y.X. Wang}
\email{wangyx@pku.edu.cn}
\affiliation{Beijing National Laboratory for Molecular Sciences, State Key Laboratory of Rare Earth Materials\\ Chemistry and Applications, 
College of Chemistry, Peking University, Beijing 100871, China}
\author{Y. Qiu}
\affiliation{NIST Center for Neutron Research, National Institute of Standards
and Technology, Gaithersburg, MD 20899, USA} 
\affiliation{Dept.\ of Materials Science and Engineering, University of Maryland, College Park, MD 20742, USA}
\author{K. Li}
\author{J.H. Lin}
\affiliation{Beijing National Laboratory for Molecular Sciences, State Key Laboratory of Rare Earth Materials\\ Chemistry and Applications, 
College of Chemistry, Peking University, Beijing 100871, China}
\author{J.R.D. Copley}
\author{R.W. Erwin}
\affiliation{NIST Center for Neutron Research, National Institute of Standards
and Technology, Gaithersburg, MD 20899, USA}
\author{B.S. Dennis}
\affiliation{Bell Laboratories, Alcatel-Lucent, Murray Hill, NJ 07974, USA}
\author{A.P. Ramirez}
\affiliation{Baskin School of Engineering, University of California, Santa Cruz, CA 95060, USA}
\affiliation{Bell Laboratories, Alcatel-Lucent, Murray Hill, NJ 07974, USA}

\date{\today}

\begin{abstract}
We report a spin $S=3/2$ triangular antiferromagnet with nearest-neighbor coupling $J=0.29$ meV in La$_2$Ca$_2$MnO$_7$. A genuinely two-dimensional, three-sublattice $\sqrt{3}\times\sqrt{3}$ order develops below
$T_N=2.80$~K~$\ll |\Theta|=25$~K. The spin excitations deviate substantially from linear spin-wave theory, suggesting that magnon breakdown occurs in the material. Such a breakdown has been anticipated in recent theoretical studies, although the excitation spectrum remains to be accounted for.
\end{abstract}


\maketitle

It is well known that quantum fluctuations renormalize N\'{e}el order only moderately in three dimensions (3D)\cite{af_kubo}, but prevent completely the long-range order even at $T=0$ for 
1D spin systems with short-range antiferromagnetic coupling\cite{ia89}.
In 2D, the quantum antiferromagnet can be either long-range ordered or remain a quantum liquid at 0 K, depending on the local geometry. The interplay of geometric frustration and quantum fluctuations
forms a frontier research area in many-body physics\cite{apr01}.

It has long been established that a 2D Ising system, which is not quantum,
has a finite-temperature transition to long-range order\cite{lo44}. When spins have continuous
symmetry, no long-range order is possible 
above zero temperature as dictated by the Mermin-Wagner theorem\cite{mw66}, 
although a critical state with power-law correlations and divergent magnetic susceptibility 
exists below the finite-temperature Kosterlitz-Thouless transition\cite{kt73}.
At zero temperature, N\'{e}el order is stable 
on a square lattice\cite{dah88}.  However, bipartite N\'{e}el order is not possible on a triangular or 
kagome lattice. This effect of geometrical frustration\cite{apr01} promises to tip the balance towards a quantum spin liquid, such as the resonating-valence-bond (RVB) state proposed by Anderson\cite{pwa74}.
On the other hand, a generalized N\'{e}el order, the three-sublattice $\sqrt{3}\times\sqrt{3}$ state (Fig.~1b), has also been proposed as the ground state\cite{jv77}. In this state, neighboring spins are oriented at 120$^o$ to each other, thereby partially relaxing the
geometrical frustration. The ground state of the 2D quantum antiferromagnet on the strongly frustrated kagome lattice is believed to be a quantum spin liquid. 
On the triangular lattice, 
the spin-half RVB and $\sqrt{3}\times\sqrt{3}$ states are so close in energy that it has taken three decades of extensive theoretical investigation for a consensus to emerge that favors the $\sqrt{3}\times\sqrt{3}$ state\cite{dah88,rps92}. 

\begin{figure*}
\includegraphics[width=1.55\columnwidth,angle=0,clip]{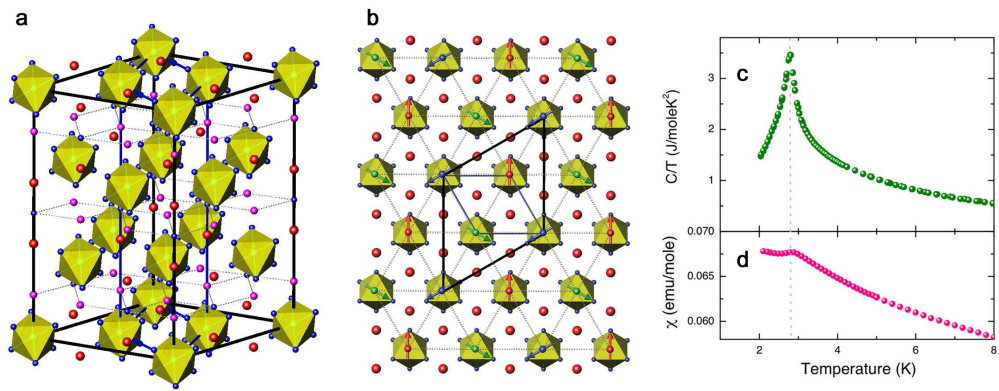}
\vskip -1em
\caption{(color online) (a) A magnetic unit cell of La$_2$Ca$_2$MnO$_7$, which contains three triangular
layers with the corner of a triangle stacking in the center of the triangles in the
neighboring layers in the close-packed $ABC$ sequence.
The red balls represent La, the pink Ca, the blue O ions. 
(b) Top view of La$_2$MnO$_6$.
The spins of the Mn$^{4+}$ ions in the centers of the MnO$_6$ octahedra form
a triangular lattice, and the arrows depict the $\sqrt{3}\times\sqrt{3}$ 
antiferromagnetic order with three color-coded spin sublattices. 
The magnetic unit cell (black lines) is three times larger than
the structural unit cell (blue lines).
(c) Specific heat
divided by temperature, as a function of temperature. (d) Magnetic susceptibility measured with a
field of 0.5 T.
}
\label{fig1}
\end{figure*}

Close to the order-disorder phase boundary of quantum spin systems, the
triangular antiferromagnet (TAF) holds a special place in the study of quantum fluctuation effects. However,
compounds that approximate the ideal situation of the theoretical model are rare. The initially
studied TAF materials are dominated by interlayer interaction, thus quasi-1D\cite{rev_tri}. Among quasi-2D materials, there was difficulty in realizing a true triangular lattice with identical nearest neighbor coupling, or without significant interaction beyond
nearest neighbors in the plane, a situation which leads to different ground states\cite{aet99}.
When the nearest-neighbor interaction dominates on a triangular lattice and leads to the correct $\sqrt{3}\!\times\!\sqrt{3}$
order in V$X_2$ ($X$ = Cl, Br or I)\cite{kh83}, $X$CrO$_2$ ($X$ = Li, Na or Cu)\cite{jls79,hk90}, $X$Fe(SO$_4$)$_2$ ($X$ = Cs or Rb)\cite{hsg99} and RbFe(MoO$_4$)$_2$\cite{les06}, such materials develop 3D order due to nonzero interplane magnetic interactions,
and 2D quantum physics is suppressed. In this work we show that the new triangular lattice compound La$_2$Ca$_2$MnO$_7$
does not suffer from any of these problems, and exhibits novel ground state behavior down to 0.04 K.

The structure of La$_2$Ca$_2$MnO$_7$ is comprised of alternate hexagonal perovskite La$_2$MnO$_6$ and
graphite-like Ca$_2$O sublayers (Fig.~1a), 
as determined by the combined neutron and x-ray diffraction study\cite{wyx04}.
The lattice parameters are $a=5.619 \AA$ and $c=17.292\AA$ at 40 mK. Magnetism is associated with the Mn$^{4+}$ ions inside the MnO$_6$ octahedra which form a perfect triangular lattice demanded by the $R\overline{3}m$ space group symmetry\cite{wyx04}. The close-packed $ABC$ stacking of triangular layers cancels possible interlayer exchange coupling due to the zero summation of the $\sqrt{3}\times\sqrt{3}$-ordered magnetic moments in a triangle (Fig.~1b). The half-filled $t_{2g}^3$ electronic configuration of Mn$^{4+}$ has spin $S=3/2$, and the magnetic susceptibility above 9 K is well described by the Rushbrooke-Wood formulas for TAF\cite{rw63} with the nearest-neighbor exchange $J=0.29$ meV and the Weiss temperature $\Theta\equiv-2JS(S+1)/k=-25$ K.   Heat capacity measurements indicate a second-order
phase transition at $T_N = 2.80$~K~$\ll |\Theta|$, which is accompanied by a small anomaly in magnetic
susceptibility (Fig.~1c,d).

To further investigate the magnetic order and excitations, neutron
scattering experiments were carried out using the Disk Chopper Time-of-Flight Spectrometer at NIST. The spectrometer simultaneously measures elastic and inelastic magnetic neutron scattering from the high purity
polycrystalline sample of 4.7 g, which was encapsulated with He exchange gas in a copper cylinder in a dilution refrigerator.
The selection of incident neutrons of 3.55 meV properly covers the spectral range of La$_2$Ca$_2$MnO$_7$ and yields an energy resolution of 0.13 meV at full-width-at-half-maximum (FWHM) for incoherent elastic scattering. A few representative spectra as a function of the wave-number $Q$ and energy $\hbar \omega$ measured from 0.04 to 30 K are shown in Fig.~2. 
\begin{figure}
\includegraphics[width=.7\columnwidth,angle=90,clip]{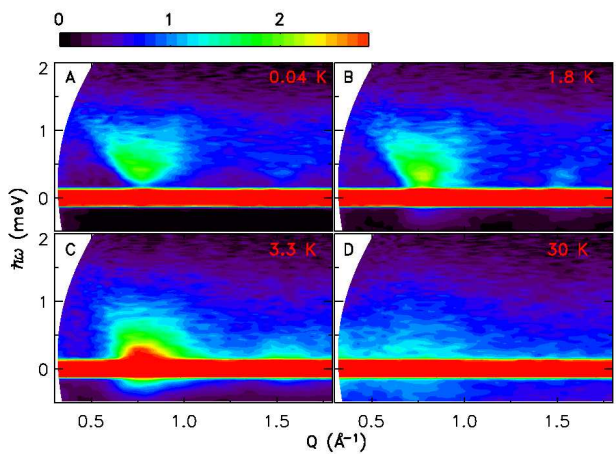}
\vskip -1em
\caption{(color online) Magnetic spectral function $S({\bf Q},\omega)$ measured at (a) 0.04 K, (b) 1.8 K, (c) 3.3 K and (d) 30 K.  The intensity in a.u.\ is indicated by the color bar at the top. 
}
\label{fig2}
\end{figure}

Let us first examine elastic magnetic neutron scattering, which is at $\hbar\omega=0$ part of the full spectrum
in Fig.~2. Raw data measured at two temperatures, well below and above $T_N$, are
shown in Supplementary Figure. While the temperature-independent 3D structural Bragg peaks
are resolution-limited, the temperature-dependent magnetic signal shows the classic, asymmetric
Warren peak profile from a 2D structure\cite{warren2D}. The integrated intensity from $Q = 0.5$ to $1\AA^{-1}$, covering
the first Warren peak, is shown as a function of temperature in the inset to Fig.~3 by red circles,
and resembles the typical order parameter of a second-order magnetic phase-transition. Extra
intensity near $T_N$ is due to the well-understood critical spin fluctuations, which peak at $T_N$. There
is negligible magnetic elastic intensity left at 30 K, which spectrum can be used as background to isolate
magnetic elastic neutron scattering intensity as shown in Fig.~3 at various temperatures.

The Warren peaks below $T_N$ in Fig.~3 testify to 2D magnetic order in La$_2$Ca$_2$MnO$_7$. 
The characteristic difference in diffraction intensities between 2D and 3D order is that the
latter produces discrete Bragg points in reciprocal space, while Bragg points coalesce into continuous
lines for 2D order. Therefore, for 2D
$\sqrt{3}\times\sqrt{3}$ order, the powder diffraction intensity is\cite{warren2D}
\begin{equation}
I(Q)=A \left(\frac{f(Q)}{\sin{2\theta}}\right)^2 \sum_i  \frac{n_i Q \left| F({\bf q}_i)\right|^2}{\sqrt{Q^2-q_i^2}}\Theta(Q-q_i),
\end{equation}
where $A$ is a constant, $f(Q)$ the atomic form factor, $1/\sin^{2}{2\theta}$ the usual Lorentz factor, $n_i$ the multiplicity of a Bragg peak, and $|F(\bf{q})|^2$ the squared magnetic structure factor at the 2D Bragg points of the $\sqrt{3}\!\times\!\sqrt{3}$ order. The unit step function
$\Theta(x)$ ensures that the Warren profile function $Q/\sqrt{Q^2-q_i^2}$ is real, which replaces the delta function in 3D
powder diffraction.
The solid line through the measured data in Fig.~3 is Eq.~(1) convoluted with the instrument resolution function. The 2D Warren peak profile describes excellently the asymmetric peaks at $q_0= 4\pi/3a=0.7455\AA^{-1}$ and $q_1\equiv 8\pi/3a=1.491\AA^{-1}$ from 40 mK to 1.8K. As a comparison, if La$_2$Ca$_2$MnO$_7$ developed a 3D magnetic order, such as that observed in CuCrO$_2$\cite{hk90}, the peak profile would have been like those indicated by the red line in Fig.~3 with discrete peaks at (1/3,1/3,$n$) and (2/3,2/3,$n$), $n$=0,1,2,3.
At 2.6 and 3.3 K, the measured spectra are broadened, which can be attributed to critical spin fluctuations in the vicinity
of $T_N$. Thus, our elastic neutron scattering results clearly demonstrate that the $\sqrt{3}\times\sqrt{3}$
order in La$_2$Ca$_2$MnO$_7$ remains 2D down to 40 mK. 

The 2D nature of the $\sqrt{3}\times\sqrt{3}$ order
in La$_2$Ca$_2$MnO$_7$ is also corroborated in the spin dynamics.
In Fig.~2d, spin excitations are noticeable well above $T_N$ at 30 K. With decreasing temperature, the
intensity grows and coalesces only at the 2D Bragg wave-vectors at $Q = q_0$ and $q_1$ (Fig.~2c). With
further cooling below $T_N$, the spectral weight of low-energy spin excitations moves progressively
to the static order, and a cone-shaped spectral distribution takes shape at $q_0$ and $q_1$ (Fig.~2a,b).
If 3D magnetic order developed in La$_2$Ca$_2$MnO$_7$, the spin excitation cones would appear
also at those extra 3D Bragg spots (1/3,1/3,$n$) and (2/3,2/3,$n$).
\begin{figure}
\includegraphics[width=.95\columnwidth,angle=0,clip]{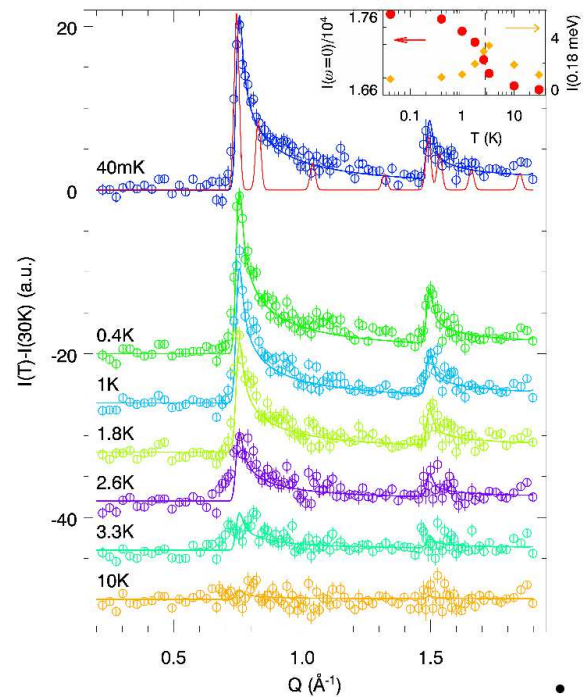}
\vskip -2em
\caption{(color online) The Warren peaks of 2D $\sqrt{3}\times\sqrt{3}$ spin order in La$_2$Ca$_2$MnO$_7$ at various temperatures. The
solid lines are resolution-convoluted fit to Eq. (1). The red line indicates the resolution-convoluted Bragg peaks when
the $\sqrt{3}\times\sqrt{3}$ order in different triangular layers form a 3D
antiferromagnetic order\cite{hk90}. The zeros of the spectra at $T \ge 0.4$ K have been shifted down for
clarity. Inset: Elastic magnetic intensity (red circles) and the intensity of critical spin fluctuations
measured at 0.18 meV and q0 (orange diamond) as a function of temperature. The dashed line
marks $T_N = 2.8$ K. 
}
\label{fig3}
\end{figure}

In previous neutron powder diffraction studies, 2D Warren magnetic peaks have been observed at finite temperature for square-lattice Ising antiferromagnets\cite{jeff_2D}, 
consistent with theoretical expectations\cite{lo44}.
Such antiferromagnetic order however is impossible on the triangular
lattice due to geometric frustration\cite{apr01}. The existence of $\sqrt{3}\times\sqrt{3}$ order in La$_2$Ca$_2$MnO$_7$ further shows that its
spins are not Ising-like: they can rotate to form the non-collinear three-sublattice order. 
The question is how can the 2D $\sqrt{3}\times\sqrt{3}$ order in La$_2$Ca$_2$MnO$_7$ have a phase transition at finite $T_N$, in apparent defiance of the Mermin-Wagner theorem?

One possibility proposed
in previous theoretical studies invokes the topological character of the $\sqrt{3}\times\sqrt{3}$ order, namely
chirality, which is absent for classic bipartite N\'{e}el order\cite{ldh84}. With its Ising-like symmetry,
chiral long-range order has been predicted theoretically for the TAF
at finite temperature\cite{ldh84}, but never before observed in a continuous-symmetry system. An open theoretical question is whether chiral long-range order would push the spin state to a long-range order at finite temperature, given that
the Kosterlitz-Thouless state at finite temperature is already power-law critical without the chirality. 
Another, more conventional, possibility is the breaking of continuous symmetry
in spin space with a finite 3-state Potts anisotropy, which is symmetry-allowed by the triple ligand
environment of Mn$^{4+}$ and  
typical ligand-field energy splittings for
octahedrally-coordinated Mn4+ are of order 10$^{-2}$ meV, much smaller than the energy resolution
of our cold neutron spectrometer. 
It is important to note
that the Potts-type order has not been previously observed in a material without 3D correlations
driving the ordering.

Now let us examine the observed spin excitations from the 2D $\sqrt{3}\times\sqrt{3}$ ground state.
The cone-shaped spectra at magnetic Bragg reflection $q_0$ in Fig.~2a indicate that
magnetic spectral weight in La$_2$Ca$_2$MnO$_7$ exists within the spin-wave cone defined by $\hbar\omega(\bf{q}) = \hbar cq$ at low temperatures. The slope of the front locus of the intensity distribution at $Q \le q_0$
measures the spin-wave velocity $c$, and it is consistent with the linear spin-wave theory prediction\cite{wz06}
$c = (3/2)^{3/2}JSa \approx 4.5$ meV$\AA$, using $J = 0.29$ meV from the Rushbrooke-Wood fit to our measured susceptibility. The consistency between the static magnetic susceptibility and the dynamic excitation spectrum indicates that La$_2$Ca$_2$MnO$_7$ indeed is an excellent material realization of the nearest-neighbor antiferromagnet on the triangular lattice.

Unlike previous studies of quasi-2D TAFs that ultimately exhibit 3D order and have excitations given by linear spin-wave theory (LSWT)\cite{sato03}, such theory completely fails to describe the the intensity distribution in La$_2$Ca$_2$MnO$_7$. The LSWT magnetic density-of-states (MDOS) at zero temperature is readily available\cite{wz06} and is shown after resolution convolution as the red curve in Fig.~4b to be compared with experimental MDOS at 40 mK obtained by integrating intensity 
from $Q = 0.35$ to 1.3$\AA^{-1}$, which includes most of the spectral weight in the first Brillouin zone (Fig.~2). This failure is anticipated by recent quantum theoretical works which show that magnons are not
elementary excitation modes at higher energy, even though the low energy cone-shaped part remains\cite{wz06}.
In general, the energy scales for the saddle point and the spin-wave bandwidth are substantially renormalized downward in the series expansions and the $1/S$ expansion studies. But our observation of the existence of magnetic excitations above the
spin-wave bandwidth $\sqrt{10} JS=1.3$ meV is not anticipated in these theories. Both multi-magnon decay and spinon-pair excitations have been proposed\cite{wz06} but the spectral function $S({\bf Q},\omega)$ 
has yet to be calculated for experimental verification. For future comparison, we provide 
spin excitation spectra at $Q = q_0$ in 
Fig.~4a and MDOS in Fig.~4b at various temperatures.
\begin{figure}
\includegraphics[width=.9\columnwidth,angle=0,clip]{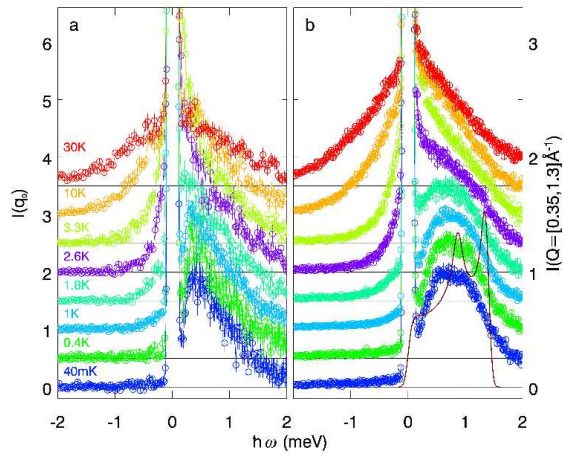}
\vskip -1em
\caption{(color online) (a) The powder energy spectra of spin excitations at the Bragg wave-number $q_0$. (b) The magnetic density-of-states $\int d{\bf Q}S({\bf Q},\omega)$. The red curve represents the
predication of linear spin-wave theory\cite{wz06}.
The same color is used in (a) and (b) to indicate the same measurement temperature.
The zeros of spectra taken at $T\ge 0.4$ K, indicated by the horizontal lines, have been shifted up for clarity.
}
\label{fig4}
\end{figure}

In summary, we report two-dimensional $\sqrt{3}\times\sqrt{3}$ spin order existing from $T_N = 2.80$ K down
to 40 mK in La$_2$Ca$_2$MnO$_7$. This allows us to investigate spin excitations from the 2D ground state
of a TAF. The spin-wave dispersion relation for low-energy spin excitations,
which survives quantum corrections, is supported by our data. The spectral weight distribution and
bandwidth of spin excitations, however, are yet to be accounted for by theory. Thus La2Ca2MnO7 represents a new model material for the study of quantum antiferromagnetism.

We thank R.R.P. Singh, Y. Ran, D.H. Lee, A. L\"{a}uchli, O. Tchernyshyov, I.A. Zaliznyak, Z. Nussinov, C. Broholm, B. Normand, O.P. Sushkov, A.V. Chubukov and R.H. McKenzie for stimulating discussions. 
Works at Peking University were supported by 
National Natural Science Foundation of China (NSFC 20571004 and 
20531010), and the DCS was partially funded by NSF (Agreement No.\ DMR-0454672).


\newpage

\renewcommand*{\figurename}{{\bf Supplementary Figure}}
\setcounter{figure}{0}

\begin{figure*}[h]
\includegraphics[width=1.55\columnwidth,angle=0,clip]{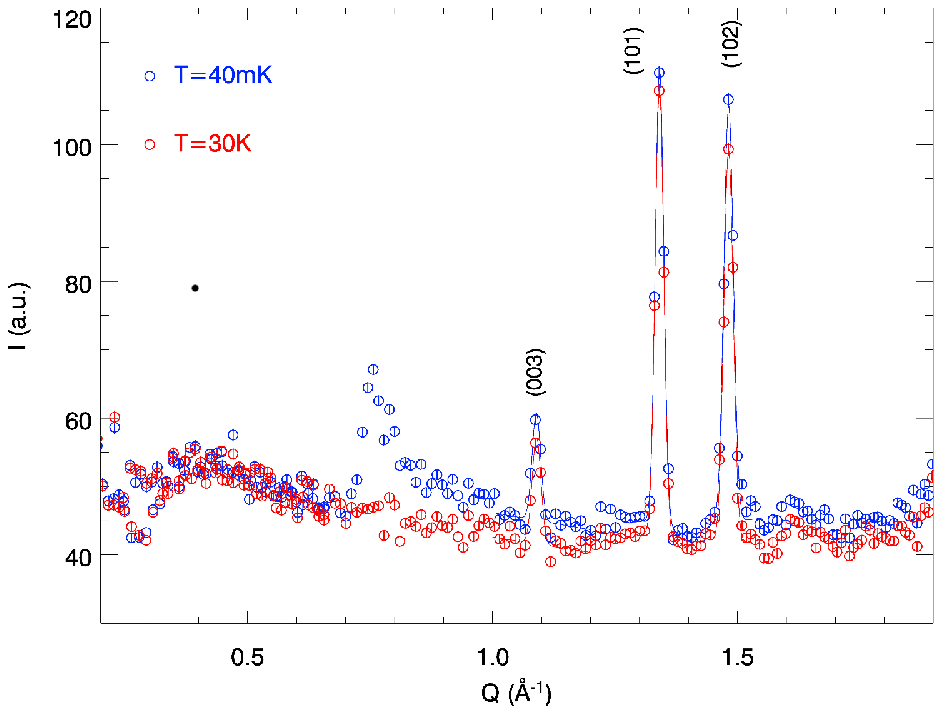}
\vskip -1em
\caption{Elastic neutron scattering spectrum of La$_2$Ca$_2$MnO$_7$ at 0.04 (blue) and 30 (red) K.
The three-dimensional structural Bragg peaks (003), (101) and (102) are resolution-limited. Their integrated intensities are temperature independent. The two-dimensional magnetic peak at $q_0=0.7455\AA^{-1}$ has the assymmetric Warren profile. The magnetic peak at $q_1=1.491\AA^{-1}$ is obscured by the (102), and is revealed in the difference plot in Fig.~3. 
}

\end{figure*}

\end{document}